\documentclass[eqsecnum,showpacs,showkeys,nofootinbib,aps]{revtex4}

\def\beq{\begin{equation}}
\def\eeq{\end{equation}}
\def\bea{\begin{eqnarray}}
\def\eea{\end{eqnarray}}
\def\nn{\nonumber}
\def\pr{\prime}

\begin{document}
\title{Intrinsic frequencies of baryons in Skyrmion theory: origin of matter wave}
\author{Soon-Tae Hong}
\email{soonhong@ewha.ac.kr} 
\affiliation{Department of Science Education and Research Institute for Basic Sciences,\\
Ewha Womans University, Seoul 120-750 Korea}
\date{\today}

\begin{abstract}
Exploiting the Hamilton quantization in the Skyrmion thoery, we investigate 
intrinsic frequencies of baryons such as nucleon and delta particles. We find 
that matter wave property of the baryons originates from these intrinsic vibrational 
modes defined on internal hypersphere.
\end{abstract}
\maketitle


It ie well known that baryons are obtainable from topological solutions, 
known as SU(2) Skymions, since the homotopy group $\Pi_{3}$(SU(2))=Z admits 
fermions~\cite{skyrme61,anw83,rab83,ws84,zb86,hong02}. Exploiting collective coordinates of isospin
rotation of the Skyrmion, Adkins et al. \cite{anw83} have performed semiclassical 
quantization to obtain static properties of baryons within 30\% of the corresponding experimental data.
Later, the SU(3) Skyrmion~\cite{su31,su32,su33,su34,su35,su36,su37,su38,su39,su310,su311,su312,su313,su314} 
has been studied to investigate its hyperfine splitting.

On the other hand, the Dirac method~\cite{di} is a well known scheme to
quantize physical systems with constraints. In this method, the Poisson
brackets in a second class constraint system are converted into Dirac
brackets to attain self consistency. Later, the Dirac Hamiltonian quantization 
has been developed to yield the improved Dirac Hamiltonian scheme~\cite{bft86,bft87,bft88,bft91,faddeev86,babelon86,harada87,niemi88,hong15} 
in which one can convert the second class constraints into first class ones. Especially, this improved Hamiltonian quantization has 
been applied to the SU(2) Skyrmion~\cite{neto97,hong99} and the SU(3) Skyrmion~\cite{hongsu300,hongsu301}.

The motivation of this paper is to revisit the standard SU(2) Skyrmion so that 
one can investigate the novel intrinsic characteristics of the baryons associated with the 
improved Dirac Hamiltonian quantization. Here we note that, even in the SU(3) Skyrmion case, the SU(2) subgroup is employed 
to describe the isospin rotational degrees of freedom. In this paper, we will not repeat the same improved Dirac 
Hamiltonian procedure for the SU(3) flavor group case.

The SU(2) Skyrmion~\cite{skyrme61,anw83,rab83,ws84,zb86,hong02}, which is a topological soliton describing fermonic baryons, is 
subject to the second class constraints~\cite{di}
$\Omega_{i}$ ($i=1,2$):
\beq
\Omega_{1}=a^{\mu}a^{\mu}-1\approx 0,~~~
\Omega_{2}=a^{\mu}\pi^{\mu}\approx 0,
\eeq
with $a^{\mu}$ and $\pi^{\mu}$ being the collective coordinates and their corresponding momenta, respectively, 
and its corresponding second class Hamiltonian is given by
\beq
H=E+\frac{1}{8{\cal I}}\pi^{\mu}\pi^{\mu},
\eeq
where $E$ and ${\cal I}$ are the static soliton energy and moment of inertia of the Skyrmion. 
Following the improved Hamitonian quantization procedure, we find 
the first class constraints $\tilde{\Omega}_{i}$ ($i=1,2$):
\beq
\tilde{\Omega}_{1}=a^{\mu}a^{\mu}-1+2\theta=0,~~~
\tilde{\Omega}_{2}=a^{\mu}\pi^{\mu}-a^{\mu}a^{\mu}\pi_{\theta}=0,
\eeq
where $(\theta,\pi_{\theta})$ are the St\"uckelberg fields. 
After tedious algebra, the first class physical fields of the original physical 
fields $a^{\mu}$ and $\pi^{\mu}$ are given by
\bea
\tilde{a}^{\mu}&=&a^{\mu}\left[1-\sum_{n=1}^{\infty}
\frac{(-1)^{n}(2n-3)!!}{n!}\frac{\theta^{n}}{a^{\nu}a^{\nu}}\right],\nn\\
\tilde{\pi}^{\mu}&=&(\pi^{\mu}-a^{\mu}\pi_{\theta})\left[1+\sum_{n=1}^{\infty}
\frac{(-1)^{n}(2n-1)!!}{n!}\frac{\theta^{n}}{a^{\nu}a^{\nu}}\right],
\eea
whose analytic forms are 
\beq
\tilde{a}^{\mu}=a^{\mu}\left(\frac{a^{\nu}a^{\nu}+2\theta}{a^{\nu}a^{\nu}}\right)^{1/2},~~~
\tilde{\pi}^{\mu}=(\pi^{\mu}-a^{\mu}\pi_{\theta})\left(\frac{a^{\nu}a^{\nu}}{a^{\nu}a^{\nu}+2\theta}\right)^{1/2}.
\eeq

We next obtain the first class Hamiltonian as follows
\beq
\tilde{H}=E+\frac{1}{8{\cal I}}(\pi^{\mu}-a^{\mu}\pi_{\theta})(\pi^{\mu}-a^{\mu}\pi_{\theta})
\frac{a^{\nu}a^{\nu}}{a^{\nu}a^{\nu}+2\theta}.
\eeq
Because the above Hamiltonian is first class, it is involutive to yield $\{\tilde{\Omega}_{i},\tilde{H}\}=0$. 
In order to find Gauss law constraint system, we introduce an equivalent first class Hamiltonian $\tilde{H}^{\pr}$:
\beq
\tilde{H}^{\pr}=\tilde{H}+\frac{1}{4{\cal I}}\pi_{\theta}\tilde{\Omega}_{2},
\label{thp}
\eeq
which thus fulfills the following identities: 
$\{\tilde{\Omega}_{1},\tilde{H}^{\pr}\}=\frac{1}{4{\cal I}}\tilde{\Omega}_{2}$ and 
$\{\tilde{\Omega}_{2},\tilde{H}^{\pr}\}=0.$ Up to now, we briefly recapitulate the Skyrmion Hamiltonian formulated in the improved Hamiltonian algorithm~\cite{hong99,hong00,hong15}.

Now, we derive the following identities,
\bea
\{\tilde{a}^{\mu},\tilde{H}\}&=&\frac{1}{4{\cal I}}\left(\tilde{\pi}^{\mu}
-\frac{\tilde{a}^{\nu}\tilde{\pi}^{\nu}}{\tilde{a}^{\nu}\tilde{a}^{\nu}}\tilde{a}^{\mu}\right),\nn\\
\{\tilde{\pi}^{\mu},\tilde{H}\}&=&\frac{1}{4{\cal I}}\frac{1}{\tilde{a}^{\nu}\tilde{a}^{\nu}}
(\tilde{a}^{\nu}\tilde{\pi}^{\nu}\tilde{\pi}^{\mu}-\tilde{\pi}^{\nu}\tilde{\pi}^{\nu}\tilde{a}^{\mu}),\nn\\
\{\tilde{a}^{\mu},\pi_{\theta}\}&=&\frac{\tilde{a}^{\mu}}{\tilde{a}^{\nu}\tilde{a}^{\nu}},\nn\\
\{\tilde{\pi}^{\mu},\pi_{\theta}\}&=&-\frac{\tilde{\pi}^{\mu}}{\tilde{a}^{\nu}\tilde{a}^{\nu}}.
\label{id4}
\eea
On the other hand, the equation of motion in the Poisson bracket form for a first class physical variable $\tilde{\Omega}$ 
governed by the Hamiltonian $\tilde{H}^{\pr}$ of interest is given by~\cite{sakurai}
\beq
\dot{\tilde{\Omega}}=\{\tilde{\Omega},\tilde{H}^{\pr}\},
\label{eomcl}
\eeq
where overdot denotes time derivative. Exploiting the identities in Eq. (\ref{id4}) and the equation of motion in Eq. 
(\ref{eomcl}), after some algebra we end up with 
\bea
\dot{\tilde{a}}^{\mu}&=&\{\tilde{a}^{\mu},\tilde{H}^{\pr}\}=\frac{1}{4{\cal I}}\tilde{\pi}^{\mu},\nn\\
\dot{\tilde{\pi}}^{\mu}&=&\{\tilde{\pi}^{\mu},\tilde{H}^{\pr}\}
=-\frac{1}{4{\cal I}}\frac{\tilde{\pi}^{\nu}\tilde{\pi}^{\nu}}{\tilde{a}^{\nu}\tilde{a}^{\nu}}\tilde{a}^{\mu}.\nn\\
\eea
We next proceed to obtain
\beq
\ddot{\tilde{a}}^{\mu}=-\frac{1}{4{\cal I}^{2}}\left[I(I+1)+\frac{1}{4}\right]\tilde{a}^{\mu},
\label{eom}
\eeq
where $I$ denotes isospin quantum number of a given baryon of interest, described by the Skyrmion. 
Here, one notes that the equations of motion for $\tilde{a}^{\mu}$ in Eq. (\ref{eom}) represent those 
for harmonic oscillators of the form
\beq
\ddot{\tilde{a}}^{\mu}=-\omega_{I}^{2}\tilde{a}^{\mu},
\label{eomharmonic}
\eeq 
where 
\beq
\omega_{I}=\frac{1}{2{\cal I}}\left[I(I+1)+\frac{1}{4}\right]^{1/2}.
\label{omegatheo}
\eeq

Next, we return to Eq. (\ref{thp}) to obtain the mass for the baryon with the isospin $I$~\cite{hong99,hong00}
\beq
M_{I}=\langle I|\tilde{H}^{\pr}|I\rangle=E+\frac{1}{2{\cal I}}\left[I(I+1)+\frac{1}{4}\right].
\eeq
For nucleon and delta particles, we find
\beq
M_{N}=E+\frac{1}{2{\cal I}},~~~
M_{\Delta}=E+\frac{2}{{\cal I}},
\eeq
respectively, and we are thus left with
\beq
E=\frac{1}{3}(4M_{N}-M_{\Delta}),~~~
{\cal I}=\frac{3}{2}(M_{\Delta}-M_{N})^{-1}.
\label{ecali}
\eeq
Inserting the experimental data~\cite{pdg} for $M_{N}$ and $M_{\Delta}$:
\beq
M_{N}^{exp}=938.9~{\rm MeV},~~~
M_{\Delta}^{exp}=1232.0~{\rm MeV},
\eeq
into Eq. (\ref{ecali}), we predict values for $E$ and ${\cal I}$ as follows, 
\beq
E=841.2~{\rm MeV},~~~
{\cal I}=(195.4~{\rm MeV})^{-1}.
\label{ecalipred}
\eeq
Now, exploiting the value for ${\cal I}$ in Eq. (\ref{ecalipred}) and the formula 
for $\omega_{I}$ in Eq. (\ref{omegatheo}), we arrive at
\beq
\omega_{N}=1.48\times 10^{23}~{\rm sec}^{-1},~~~
\omega_{\Delta}=2.97\times 10^{23}~{\rm sec}^{-1}.
\label{omegandelta}
\eeq
These are our main results.

Now, it seems appropriate to address couple of comments on the intrinsic frequencies of the baryons.
During the baryons rotate via collective motion, inside the Skyrmion there exist vibration modes 
characterized by the intrinsic vibration frequencies in Eq. (\ref{omegandelta}). 
The vibrational motion seems to be similar to heartbeats of human being. 
Moreover, we arrive at the novel aspect that matter wave property is due to these intrinsic vibrational modes. 

Next, the intrinsic frequencies and intrinsic masses 
depend on the isospin quantum number $I$ so that the quantum number $I$ can characterize 
species of the baryons like their fingerprints. In other words, the intrinsic frequencies 
in Eq. (\ref{omegandelta}) specify the masses of the baryons. We thus find that the 
fermionic particle masses are closely related to their intrinsic frequencies in the Skyrmion theory. 
It is also interesting to note that the collective coordinates $\tilde{a}^{\mu}$ are 
defined on $S^{3}$ manifold to create the oscillation on this hypersphere.

\end{document}